\documentclass[journal,10pt,twoside]{IEEEtran}
\usepackage{cite}
\usepackage{graphicx}
\usepackage{psfrag}
\usepackage{url}
\usepackage{amsmath}
\usepackage{array}
\usepackage{amssymb}
\usepackage{mathtools}
\usepackage{amsfonts}
\usepackage{graphicx}
\usepackage{epstopdf}
\usepackage{algorithm}
\usepackage{algorithmic}

\usepackage{setspace}
\usepackage{amscd}
\usepackage{mathrsfs}
\usepackage{epsfig}
\usepackage{color}
\usepackage{textcomp}
\usepackage{tikz}
\usepackage{verbatim}
\usetikzlibrary{shapes,snakes,plotmarks}

\usepackage{caption}
\usepackage{subcaption}

\newcommand{\boundellipse}[3]{(#1) ellipse (#2 and #3)}

\DeclareMathOperator{\diag}{diag}
\DeclareMathOperator{\trace}{Tr}
\DeclareMathOperator*{\argmax}{argmax}
\DeclareMathOperator*{\argmin}{argmin}
\DeclareMathOperator*{\indmax}{indmax}

\makeatletter
\newcommand\fs@spaceruled{\def\@fs@cfont{\bfseries}\let\@fs@capt\floatc@ruled
  \def\@fs@pre{\vspace{0.5\baselineskip}\hrule height.7pt depth0pt \kern2pt}%
  \def\@fs@post{\kern2pt\hrule\relax}%
  \def\@fs@mid{\kern2pt\hrule\kern2pt}%
  \let\@fs@iftopcapt\iftrue}
\makeatother

\title{Clustering Based Activity Detection Algorithms for Grant-Free Random Access in Cell-Free Massive MIMO}

\begin{document}
	\author{Unnikrishnan Kunnath Ganesan, \IEEEmembership{Graduate Student Member, IEEE,} Emil Bj\"ornson, \IEEEmembership{Senior Member, IEEE} and \\Erik G. Larsson, \IEEEmembership{Fellow, IEEE}%
		
		\thanks{Unnikrishnan Kunnath Ganesan and Erik G. Larsson were supported in part by ELLIIT and in part by Swedish Research Council
			(VR). Emil Björnson was supported by the Grant 2019-05068 from the Swedish Research Council. This article was presented at 21st IEEE International Workshop on Signal Processing Advances in Wireless Communications (SPAWC 2020)\cite{ganesan2020algorithm}. }

		\thanks{Unnikrishnan Kunnath Ganesan and Erik G. Larsson are with the Department of Electrical Engineering (ISY), Linköping University,	581 83 Linköping, Sweden (e-mail: unnikrishnan.kunnath.ganesan@liu.se;erik.g.larsson@liu.se).}

		\thanks{Emil Björnson is with the Department of Electrical Engineering (ISY), Linköping University, 581 83 Linköping, Sweden, and is also with the KTH Royal Institute of Technology, 114 28 Stockholm, Sweden (e-mail: emil.bjornson@liu.se).}

	}

\markboth{IEEE Transactions on Communications, Vol. XX, No. XX, 202X}{Kunnath Ganesan \lowercase{\textit{et al.}}: Clustering Based Activity Detection Algorithms}

\maketitle
\thispagestyle{empty}

\begin{abstract}
Future wireless networks need to support massive machine type communication (mMTC) where a massive number of devices accesses the network and massive MIMO is a promising enabling technology. Massive access schemes have been studied for co-located massive MIMO arrays. In this paper, we investigate the activity detection in grant-free random access for mMTC in cell-free massive MIMO networks using distributed arrays. Each active device transmits a non-orthogonal pilot sequence to the access points (APs) and the APs send the received signals to a central processing unit (CPU) for joint activity detection. The maximum likelihood device activity detection problem is formulated and algorithms for activity detection in cell-free massive MIMO are provided to solve it. The simulation results show that the macro diversity gain provided by the cell-free architecture improves the activity detection performance compared to co-located architecture when the coverage area is large. 
\end{abstract}

\begin{IEEEkeywords} 
Activity Detection, Grant-Free Random Access, Cell-Free massive MIMO, massive machine-type communications (mMTC), Internet-of-Things (IoT).
\end{IEEEkeywords}

\section{Introduction}	
\IEEEPARstart{T}{he} data traffic in wireless networks has grown tremendously in the last decade. There is a growing consensus that the future wireless networks should support three generic services namely enhanced mobile broadband (eMBB), massive machine type communications (mMTC) and ultra reliable low latency communications (URLLC). URLLC and mMTC~\cite{dutkiewicz2017massive,bockelmann2016massive} are two key features of Internet-of-Things (IoT) services envisioned in 5G and beyond 5G wireless networks~\cite{bockelmann2016massive,bana2019massive}. URLLC is a critical IoT service for time-critical communications which enables ultra-high reliability and/or ultralow latency at a variety of data rates. MMTC in IoT is meant to support a massive number of low-cost devices which carry very small data packets and require extreme coverage. Massive MIMO for IoT connectivity is still a developing topic and in this paper, we study the cell-free massive MIMO networks for mMTC services. One of the main challenges of mMTC is that the network should be able to support a large number of devices over the same time and frequency resources while keeping the battery lives of the devices as long as possible. Massive multiple input multiple output (MIMO) was shown to be a promising technology to support massive access in~\cite{marzetta2010noncooperative,bana2019massive} by exploiting the spatial degrees of freedom available in the network to let many users transmit simultaneously. 

Conventional grant-based massive random access schemes are studied in \cite{pratas2012code, sorensen2014massive, bjornson2016random}. In the grant-based approach, each active device randomly picks a pilot sequence from a shared pool of orthogonal sequences, and uses the selected sequence to inform the base station that it has data to transmit. The base station needs to resolve the collisions that occur and a grant of resources will be provided to selected devices based on collision resolution. In wireless systems, the channel coherence interval is limited and hence, the set of orthogonal pilot sequences is finite. Grant-based protocols permit simple signal processing at the base station. A key feature of mMTC is that the traffic is sporadic with high risk of collision of potentially active users and with very small payloads.  Thus, in the massive random access scenario, the probability of multiple active devices selecting the same sequence is quite high. Thus, the grant-based random access protocols suffer from access failure due to collisions, and hence, increase the average latency. Moreover, to resolve the collisions, the signaling overhead is quite large compared to the short payload each device has to send in mMTC applications. Authors in~\cite{yang2016neighbor} propose a neighbor-aware multiple access protocol that improves system throughput by exploiting the broadcasted acknowledgment signals in the network. However, considering the average latency incurred and short payloads, it is inefficient to use conventional grant-based access methods for mMTC. 

To overcome the limitations of grant-based random access schemes, various grant-free protocols~\cite{shahab2020grant} have been proposed for the active devices to access the wireless network without a grant. The grant-free approach reduces the access latency and signaling overhead compared to grant-based approaches at the expense of sophisticated signal processing at the base station. In the grant-free random access scheme, each device is assigned a unique pilot sequence and the active devices access the network using this sequence. To enable data transmission for mMTC, channel estimates are required for which identification of active users is crucial. Thus, in the paper, we focus on the activity detection problem in a grant-free random access scheme. To support a massive number of devices in mMTC and due to the limited channel coherence interval length, assigning orthogonal pilot sequences to each device is not feasible. Hence, the users are assigned with unique non-orthogonal pilot sequences and hence the received signal at the base station suffer from severe co-channel interference. Thus, the activity detection is a challenging problem in the grant-free massive random access schemes. The sparse nature of the device activity pattern helps to formulate the activity detection problem as a compressive sensing (CS) problem~\cite{bockelmann2013compressive,monsees2015compressive} and greedy pursuit algorithms were leveraged to solve it~\cite{gao2015compressive,du2017efficient,xiao2019grant}, with the assumption that the number of active users is known a priori. Bayesian inference based algorithms like approximate message passing (AMP) are utilized in~\cite{liu2018massive,liu2018sparse,senel2017device,senel2018grant,chen2019multi} which are computationally efficient for activity detection. However, the performance of CS based algorithms degrade severely when the number of active devices is larger than the pilot sequence length~\cite{donoho2009message,senel2017device}. Random and structured sparsity learning based multi-user detection was studied in \cite{ding2019sparsity}. Deep learning based activity detection approaches are considered in~\cite{bai2019deep,zhang2019dnn}.

A covariance-based approach is proposed in \cite{haghighatshoar2018improved} for device activity detection which performs better than the existing CS based schemes. The covariance based approach overcomes the limitation in traditional CS based techniques where the number of active devices are required to be  smaller than the pilot sequence length. Joint activity and data detection using a covariance-based approach is proposed in \cite{chen2019covariance} and the performance analysis of the covariance-based approach is provided showing the superiority over the AMP based approach by exploiting the asymptotic properties of maximum likelihood estimator. The covariance based method makes better use of the multiple antennas to improve detection accuracy compared to AMP based CS approaches. The activity detection in unsourced random access \cite{polyanskiy2017perspective} where all devices use the same codebook is studied in \cite{fengler2019grant} and is shown to have high spectral efficiencies when a covariance based recovery algorithm is employed at the receiver. 

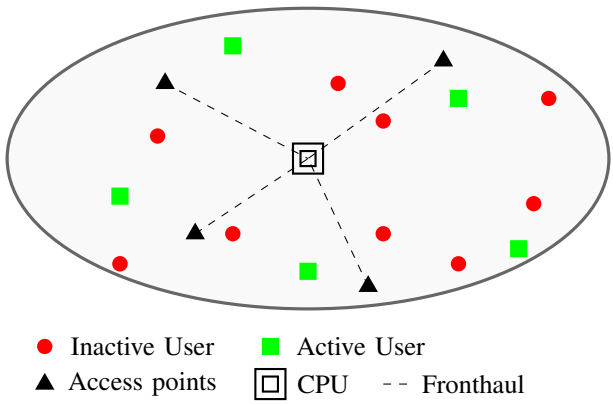
\begin{figure}[t]
	\centering
	\begin{tikzpicture}
		\filldraw[color=black!60, fill=gray!5, very thick](0,0) \boundellipse{0,0}{4}{2};
		\filldraw[red, thick] (-1,-1) circle (2.5pt);
		\filldraw[red, thick] (1,0.5) circle (2.5pt);
		\filldraw[red, thick] (-2,0.3) circle (2.5pt);
		\filldraw[red, thick] (0.4,1) circle (2.5pt);
		\filldraw[red, thick] (1,-1) circle (2.5pt);
		\filldraw[red, thick] (2,-1.4) circle (2.5pt);
		\filldraw[red, thick] (-2.5,-1.4) circle (2.53pt);
		\filldraw[red, thick] (3,-0.6) circle (2.5pt);
		\filldraw[red, thick] (3.2,0.8) circle (2.5pt);
		
		\filldraw[green, thick] (-2.6,-0.6) rectangle (-2.4,-0.4); 
		\filldraw[green, thick] (1.9,0.7) rectangle (2.1,0.9); 
		\filldraw[green, thick] (-0.1,-1.4) rectangle (0.1,-1.6); 
		\filldraw[green, thick] (-1.1,1.4)  rectangle (-0.9,1.6); 
		\filldraw[green, thick] (2.7,-1.3)  rectangle (2.9,-1.1); 
		
		\node[fill=black,regular polygon, regular polygon sides=3,inner sep=1.5pt] at (-1.9,1) {};
		\draw[-, dashed] (-1.9,1) -- (0,0);
		\node[fill=black,regular polygon, regular polygon sides=3,inner sep=1.5pt] at (-1.5,-1) {};	
		\draw[-, dashed] (-1.5,-1) -- (0,0);
		\node[fill=black,regular polygon, regular polygon sides=3,inner sep=1.5pt] at (1.8,1.3) {};
		\draw[-, dashed] (1.8,1.3) -- (0,0);
		\node[fill=black,regular polygon, regular polygon sides=3,inner sep=1.5pt] at (0.8,-1.7) {};
		\draw[-, dashed] (0.8,-1.7) -- (0,0);	
		
		\draw[black, thick] (-0.2,-0.2) rectangle (0.2,0.2);
		\draw[black, thick] (-0.1,-0.1) rectangle (0.1,0.1);
		
		\filldraw[red, thick] (-3.5,-2.5) circle (2.5pt);
		\node at (-2.2,-2.5) {Inactive User};
		\filldraw[green, thick] (-0.6,-2.6) rectangle (-0.4,-2.4); 
		\node at (0.7,-2.5) {Active User};
		\node[fill=black,regular polygon, regular polygon sides=3,inner sep=1.5pt] at (-3.5,-3) {};	
		\node at (-2.2,-3) {Access points};
		\draw[black, thick] (-0.7,-3.2) rectangle (-0.3,-2.8);
		\draw[black, thick] (-0.6,-3.1) rectangle (-0.4,-2.9);
		\node at (0.2,-3) {CPU};
		
		\draw[-, dashed] (1,-3) -- (1.4,-3);
		\node at (2.2,-3) {Fronthaul};
	\end{tikzpicture}
	\caption{Cell-Free Network Model for mMTC.}
	\label{fig:network}
	\vspace{-15pt}
\end{figure}

To provide a uniform and high per-user data rates in future wireless networks, densification of the network infrastructure by increasing the number of antennas per cell and deploying many access points (APs), is considered. However, network densification increases the inter-cell interference. Towards this, cell-free massive MIMO was proposed wherein the concept of cell is removed and all device communicates with all the APs. In the cell-free architecture, all the APs are connected to a central processing unit (CPU) through a fronthaul as shown in Fig.~\ref{fig:network}. The CPU is responsible for jointly processing data and serving all the users in the network. Cell-free massive MIMO was shown to be a promising approach to overcome inter-cell interference limitation posed by network densification~\cite{ngo2017cell,interdonato2019ubiquitous}. In cell-free architecture, the path loss from a device to different APs differ by many orders of magnitude. In~\cite{wang2020grant}, grant-free access in cell-free architecture is studied with the assumption of ideal favorable propagation in massive MIMO systems~\cite{marzetta2016fundamentals,chen2018channel}. Favorable propagation in cell-free networks is more likely to occur when more APs are in the systems and the users are spatially well-separated, which cannot be guaranteed in general, and ideal favorable propagation is unlikely.

\subsection{Contributions} \label{sec:contributions}

In this paper, we investigate and develop algorithms for activity detection for grant-free random access in cell-free massive MIMO networks without any assumptions on the network geometry leveraging the covariance-based approach. Most of the existing literature studies random access in a co-located scenario in single and multi-cell environments~\cite{xu2015active,chen2021sparse}. In \cite[Sec. 3.5]{fengler2021sparse}, Fengler proposes an activity detection algorithm for cell-free setup. In Fengler's algorithm, each AP decodes the activity pattern of all users and shares their estimates with each other in every iteration. After each iteration, thresholding is done considering the estimates from all APs and the devices are marked as inactive which did not meet threshold criteria by at least one AP. With subsequent iterations, the false alarm rate improves but the missed detection rate does not improve much.
In our work, we consider a joint activity detection at the CPU, which improves the activity detection performance in terms of false alarm and missed detection rates. Cell-free MIMO networks can provide better coverage for mMTC due to the shorter propagation distances and are more robust to the shadow fading effects compared to co-located MIMO at the expense of the increased cost of the fronthaul infrastructure. We show that with sophisticated signal processing at the CPU, cell-free networks can obtain better activity detection performance compared to co-located MIMO networks. This reduces the energy consumption of mMTC devices as the fraction of time the device needs to be active becomes shorter. Thereby, keeping the battery life of mMTC devices as long as possible.

For activity detection in a cell-free network, the optimal method would be, if all APs are contributing to the activity detection for all users. But this is unnecessarily computationally complex. Since only a few APs are close to each user, we consider a cluster of APs with good channels to the user. In~\cite{ganesan2020algorithm}, we considered a special case where the cluster size was 1. In this paper, we develop algorithms to improve the  activity detection performance by considering the information from a cluster of dominant APs. Thus, we propose novel clustering-based activity detection algorithms utilizing information from a cluster of dominant APs for activity detection, and simulations show that the performance improves with the clustering-approach. 

The main contributions of this paper can be summarized as follows:
\begin{itemize}
\item We investigate the grant-free massive random access in cell-free massive MIMO networks and formulate the maximum likelihood activity detection problem leveraging the covariance-based approach.

\item We study the SNR achieved at the APs for cell-free massive MIMO networks in different deployment scenarios considering shadow fading, dense deployment of APs, and cell area for power limited mMTC devices. We show that the outage probability is less in cell-free networks compared to co-located networks. Hence, cell-free networks provide broad coverage and enable IoT services for many devices.

\item We propose an algorithm for device activity detection based on the single dominant AP for each device for cell-free massive MIMO networks. This algorithm was presented in the conference version \cite{ganesan2020algorithm}. 

\item We propose a novel clustering-based activity detection algorithm for activity detection which improves the activity detection performance with the cost of sophisticated signal processing at the CPU. 

\item We quantify the complexity involved in the detection process and by introducing clustering, we can identify a suitable trade-off between the performance and complexity.

\item  We propose a novel algorithm which reduces the computational time required for the activity detection, based on clustering and parallelism by exploiting the sparse nature of activity pattern. 

\item Our simulations show that the cell-free massive MIMO networks can provide better activity detection performance in an mMTC scenario compared to a co-located network. They also show that activity detection performance can be further improved by clustering-approach. 

\item We study the impact of a capacity-limited fronthaul on the activity detection performance and show that we can achieve a performance similar to lossless fronthaul with contemporary fronthaul technology. 

\end{itemize}

The organization of the rest of the paper is as follows. The signal model and the activity detection problem formulation is explained in Sec. \ref{sec:SignalModel}. The activity detection algorithms are provided in Sec. \ref{sec:DeviceActivityDetectionAlgorithm}. Sec. \ref{sec:SimulationResults} provides the simulation results which shows the performance of the proposed algorithms and Sec. \ref{sec:Conclusion} provides the concluding remarks. 	

\textit{\textbf{Reproducible research:}} All the simulation results can be reproduced using the Matlab code available at:\\ \url{https://github.com/emilbjornson/grant-free}. 

\textbf{\textit{Notations:}} Bold, lowercase letters are used to denote vectors and bold, uppercase letters are used to denote matrices. $\Re(\cdot)$ and $\Im(\cdot)$ denotes the real and imaginary parts, respectively. $\mathbb{R}$ and $\mathbb{C}$ denote the set of real and complex numbers respectively. The operations $(\cdot)^\text{T}$ and  $(\cdot)^\text{H}$ denote transpose and Hermitian transpose, respectively. $\mathcal{CN}(0,\sigma^2)$ denotes a circularly symmetric complex Gaussian random variable with zero mean and variance equal to $\sigma^2$. $\mathbf{I}_N$ and $\mathbf{0}_N$ represent the $N \times N$ identity matrix and $N\times1$ zero vector, respectively. The operation $|\cdot|$ denotes the determinant. The operation $|\cdot|_\text{c}$ denotes the cardinality of set. $ \diag (\mathbf{a}) $ represents a diagonal matrix with diagonal entries are elements from $\mathbf{a}$.

\section{Signal Model And Problem Formulation}
\label{sec:SignalModel}

Consider a cell-free massive MIMO wireless network as illustrated in Fig. \ref{fig:network} with $M$ arbitrarily distributed APs each equipped with $N$ antennas and serving $K$ arbitrarily distributed single antenna users. All the $M$ APs are assumed to be connected to a CPU through a lossless infinite capacity fronthaul. Due to the sporadic nature of the traffic in the massive access scenario of mMTC, only a small fraction of the $K$ users are active at any given time instant. In this paper, we assume that each device transmits independently with an activation probability $\epsilon\ll 1$. Let $a_k\in\{0,1\}$  where $a_k =1$ denotes that the $k^{th}$ device is active and $a_k = 0$ that it is inactive and $\text{Pr}(a_k=1)=\epsilon$ and $\text{Pr}(a_k=0)=1-\epsilon$. Let $\mathbf{a}=(a_1,a_2,\cdots,a_K)$ denote the activity of $ K $ users at any time instant. Due to the sporadic nature of mMTC traffic, the vector $\mathbf{a}$ will be sparse. The set of active users is denoted by $\mathcal{A}$ i.e., $\mathcal{A} = \{k : a_k=1\}$.  

We consider that all the APs and the devices are geographically separated from each other and hence the channels between the devices and the APs can be considered independent. Moreover, we consider that the $ N $ antennas at each AP are sufficiently separated to have independent fading between them. Hence, the channel gain between the $n^{th}$ antenna in the $m^{th}$ AP to device $k$ is given by 
\begin{equation}
g_{mnk} = \beta_{mk}^{\frac{1}{2}}h_{mnk}
\end{equation}
where $\beta_{mk}$ is the large-scale fading coefficient between the $m^{th}$ AP and the user $k$ and $h_{mnk} \sim \mathcal{CN}(0,1)$ is the small-scale fading coefficient. We assume that the large-scale fading coefficient parameters $\{\beta_{mk}\}$ are known at the CPU \cite{wang2018fly,fengler2019non}. Throughout this paper, we consider a block fading scenario where each channel remains constant during a coherence interval \cite[Ch.2]{marzetta2016fundamentals} and all the channels are independently distributed. Let $\tau_c$ be the number of channel uses per coherence interval. Due to the large number of users, typically $K\gg \tau_c$, assigning orthogonal pilot sequences to each user is not feasible. Instead we assign non-orthogonal unique signature sequence, $\mathbf{s}_k\in \mathbb{C}^{L\times1}$ to each user $k$, where $L\leq\tau_c$ is the signature sequence length. We assume that the signature sequences of all the users are known at the CPU. Moreover, we assume that all the devices are synchronized during the transmission, which means in an orthogonal frequency division multiplexing system, the time delays from the different devices are well within the cyclic prefix. 

The signal $\mathbf{y}_{mn}\in \mathbb{C}^{L\times 1}$ received at the $n^{th}$ antenna of the $m^{th}$ AP is given by 
\begin{equation}
\begin{aligned}
\mathbf{y}_{mn} & = \sum_{k=1}^{K}a_k\rho_k^{\frac{1}{2}} g_{mnk} \mathbf{s}_k + \mathbf{w}_{mn} \\
& = \mathbf{SD_aD}_{\boldsymbol{\rho}}^{\frac{1}{2}}\mathbf{g}_{mn}  + \mathbf{w}_{mn},
\end{aligned}
\end{equation}
where $\mathbf{S}=[\mathbf{s}_1 \ \mathbf{s}_2 \ \dots \ \mathbf{s}_K] \in \mathbb{C}^{L\times K}$ is the collection of all signature sequences, $\rho_k$ is the power transmitted by user $k$, $ \mathbf{D_a} = \diag(\mathbf{a})$, $ \mathbf{D}_{\boldsymbol{\rho}} = \diag(\rho_1,\rho_2,\dots,\rho_K)$, $\mathbf{g}_{mn} = [g_{mn1} \ g_{mn2} \ \dots \ g_{mnK}]^{\text{T}} \in \mathbb{C}^{K\times1}$ is the channel vector from all $K$ users to the $n^{th}$ antenna of the $m^{th}$ AP and $\mathbf{w}_{mn}\sim \mathcal{CN}(\mathbf{0},\sigma^2\mathbf{I}_L)$ is the independent additive white Gaussian noise vector.

Thus, the signal $\mathbf{Y}_m\in\mathbb{C}^{L\times N}$ received at the $m^{th}$ AP can be expressed as 
\begin{equation}
\label{eqn:Ym_BS}
\mathbf{Y}_m = \mathbf{SD_aD}_{\boldsymbol{\rho}}^{\frac{1}{2}}\mathbf{G}_{m}  + \mathbf{W}_{m},
\end{equation}
where $\mathbf{G}_m = [\mathbf{g}_{m1} \ \mathbf{g}_{m2} \ \dots \ \mathbf{g}_{mN}]\in \mathbb{C}^{K\times N}$ is the channel matrix between the $K$ users and the $m^{th}$ AP and $\mathbf{W}_{m} = [\mathbf{w}_{m1} \ \mathbf{w}_{m2} \ \dots \ \mathbf{w}_{mN}] \in \mathbb{C}^{L\times N}$ is the noise matrix. 

Let the collection of signals be
\begin{equation}
\scriptsize
\label{eqn:Y_CPU}
\begin{aligned}
\mathbf{Y} & = \left[ 
\begin{matrix}
\mathbf{Y}_1 \\ \mathbf{Y}_2  \\ \vdots \\ \mathbf{Y}_M
\end{matrix}
\right]  = 
\left[ 
\begin{matrix}
\mathbf{SD_aD}_{\boldsymbol{\rho}}^\frac{1}{2}\mathbf{G}_1 \\ \mathbf{SD_aD}_{\boldsymbol{\rho}}^\frac{1}{2}\mathbf{G}_2  \\ \vdots \\ \mathbf{SD_aD}_{\boldsymbol{\rho}}^\frac{1}{2}\mathbf{G}_M 
\end{matrix}
\right] + \mathbf{W}
\\
& = 
\left[  \begin{matrix}
\mathbf{S} & \mathbf{0} & \dots & \mathbf{0} \\
\mathbf{0} & \mathbf{S} & \dots & \mathbf{0} \\
\vdots & \vdots & \ddots & \vdots \\
\mathbf{0} & \mathbf{0} & \dots & \mathbf{S} 
\end{matrix}
\right]
\left[  \begin{matrix}
\mathbf{D_aD}_{\boldsymbol{\rho}}^\frac{1}{2} & \mathbf{0} & \dots & \mathbf{0} \\
\mathbf{0} & \mathbf{D_aD}_{\boldsymbol{\rho}}^\frac{1}{2} & \dots & \mathbf{0} \\
\vdots & \vdots & \ddots & \vdots \\
\mathbf{0} & \mathbf{0} & \dots & \mathbf{D_aD}_{\boldsymbol{\rho}}^\frac{1}{2} 
\end{matrix}
\right]
\left[ 
\begin{matrix}
\mathbf{G}_1 \\ \mathbf{G}_2  \\ \vdots \\ \mathbf{G}_M
\end{matrix}
\right] +  \mathbf{W}, 
\end{aligned}
\end{equation} 
where $\mathbf{W} = [\mathbf{W}_1^{\text{T}} \ \mathbf{W}_2^{\text{T}} \ \dots \ \mathbf{W}_M^{\text{T}} ]^{\text{T}}$. From  (\ref{eqn:Y_CPU}), it can be seen that the columns of $\mathbf{Y}$ are independent and each column is distributed as $\mathbf{Y}(:,i) \sim \mathcal{CN}(\mathbf{0}_{LM},\mathbf{Q})$, $\forall i=1,2,\dots,N$, where $\mathbf{Q}$ is the covariance matrix given by 
\begin{equation}
\small
\mathbf{Q} = 
\left[  \begin{matrix}
\mathbf{SD}_{\boldsymbol{\gamma}}\mathbf{D}_{\boldsymbol{\beta}_1}\mathbf{S}^{\text{H}} & \mathbf{0}_L & \dots & \mathbf{0}_L \\
\mathbf{0}_L & \mathbf{SD}_{\boldsymbol{\gamma}}\mathbf{D}_{\boldsymbol{\beta}_2}\mathbf{S}^{\text{H}} & \dots & \mathbf{0}_L \\
\vdots & \vdots & \ddots & \vdots \\
\mathbf{0}_L & \mathbf{0}_L & \dots & \mathbf{SD}_{\boldsymbol{\gamma}}\mathbf{D}_{\boldsymbol{\beta}_M}\mathbf{S}^{\text{H}}
\end{matrix}
\right] + \sigma^2\mathbf{I}_{LM},
\end{equation}
where $\mathbf{D}_{\boldsymbol{\beta}_m}$ is a diagonal matrix with diagonal elements corresponding to the large-scale fading coefficient from $K$ users to $m^{th}$ AP, i.e., $\mathbf{D}_{\boldsymbol{\beta}_m} = \diag(\boldsymbol{\beta}_m)$ where $\boldsymbol{\beta}_m=(\beta_{m1},\beta_{m2},\dots,\beta_{mK})$ and $\mathbf{D}_{\boldsymbol{\gamma}} = \diag(\boldsymbol{\gamma})$, where $\boldsymbol{\gamma} = (a_1\rho_1,a_2\rho_2,\dots,a_K\rho_K)$. 

By utilizing the block-diagonal structure of the covariance matrix $\mathbf{Q}$, the likelihood of $\mathbf{Y}$ given $\boldsymbol{\gamma}$ is 
\begin{equation}
\begin{aligned}
p(\mathbf{Y}|\boldsymbol{\gamma}) & = \prod_{m=1}^{M}\prod_{n=1}^{N} \frac{1}{|\pi\mathbf{Q}_m|}\exp\left( -\mathbf{y}_{mn}^{\text{H}} \mathbf{Q}_m^{-1} \mathbf{y}_{mn} \right) \\
& = \prod_{m=1}^{M} \frac{1}{|\pi\mathbf{Q}_m|^N}\exp(-\trace(\mathbf{Q}_m^{-1}\mathbf{Y}_m\mathbf{Y}_m^{\text{H}})),
\end{aligned}
\end{equation}
where $ \mathbf{Q}_m = \mathbf{SD}_{\boldsymbol{\gamma}}  \mathbf{D}_{\boldsymbol{\beta}_m}\mathbf{S}^{\text{H}} + \sigma^2\mathbf{I}_L$. The maximum likelihood estimate of $\boldsymbol{\gamma}$ can be found by maximizing $p(\mathbf{Y}|\boldsymbol{\gamma})$ or equivalently minimizing $-\log(p(\mathbf{Y}|\boldsymbol{\gamma}))$ which is given by 
\begin{equation}
\label{eqn:ML_func}
\begin{aligned}
\boldsymbol{\gamma}^* = & \ \underset{\boldsymbol{\gamma}}{\arg\min} \sum_{m=1}^{M} \log|\mathbf{Q}_m| + \trace\left(\mathbf{Q}_m^{-1}\frac{\mathbf{Y}_m\mathbf{Y}_m^{\text{H}}}{N}\right) \\
& \text{subject to } \boldsymbol{\gamma} \geq \mathbf{0}_K. 
\end{aligned}
\end{equation}

To perform the activity detection, all the received signals at the APs need to be passed to the CPU for $L\geq N$. When $L<N$, AP $m$ sends the sample covariance $\mathbf{Y}_m\mathbf{Y}_m^{\text{H}}$ to the CPU to reduce the fronthaul usage. The CPU needs to solve the optimization problem in (\ref{eqn:ML_func}). For $M=1$, the co-located architecture case, a covariance-based coordinate descent algorithm is proposed in \cite{haghighatshoar2018improved} for device activity detection. However, for a cell-free architecture, due to the presence of $M>1$ summation terms in (\ref{eqn:ML_func}), the brute force approach to solve (\ref{eqn:ML_func}) requires huge complexity and the complexity increases exponentially with $ M $. In Sec. \ref{sec:DeviceActivityDetectionAlgorithm} we develop algorithms for the device activity detection that has affordable complexity, while making use of information obtained at multiple APs.

\section{Device Activity Detection}
\label{sec:DeviceActivityDetectionAlgorithm}
In this section, we study the cost function (\ref{eqn:ML_func}) and exploit the features of cell-free architecture to develop algorithms for activity detection in grant-free random access schemes. 

\subsection{Coordinate Descent Cost Function}\label{sec:CoordinateDescentCostFunction}

Let 
\begin{equation}
f(\boldsymbol{\gamma}) = \sum_{m=1}^{M} \log|\mathbf{Q}_m| + \trace\left(\mathbf{Q}_m^{-1}\frac{\mathbf{Y}_m\mathbf{Y}_m^{\text{H}}}{N}\right)
\label{eqn:ML_costfunc}
\end{equation}
be the cost function which needs to be minimized in (\ref{eqn:ML_func}). Define 
\begin{equation}
	f^m(\boldsymbol{\gamma}) = \log|\mathbf{Q}_m| + \trace\left(\mathbf{Q}_m^{-1}\frac{\mathbf{Y}_m\mathbf{Y}_m^{\text{H}}}{N}\right)
\end{equation} 
be the cost function  associated with the $m^{th}$ block in (\ref{eqn:ML_costfunc}). Then we can write $f(\boldsymbol{\gamma}) = \sum_{m=1}^{M}f^m(\boldsymbol{\gamma})$. Setting $\mathbf{Q}_m$ as a function of ${\boldsymbol{\gamma}}$, i.e., 
\begin{align}
\mathbf{Q}_m(\boldsymbol{\gamma}) & = \mathbf{S}\mathbf{D}_{\boldsymbol{\gamma}}\mathbf{D}_{\boldsymbol{\beta}_m}\mathbf{S}^{\text{H}} + \sigma^2\mathbf{I}_L \\
\label{eqn:RankOneUpdates}
&= \sum_{k=1}^{K}\gamma_k\beta_{mk}\mathbf{s}_k\mathbf{s}_k^{\text{H}} +  \sigma^2\mathbf{I}_L,
\end{align}
we can see $\mathbf{Q}_m$ as a sum of $K$ rank-one updates to $\sigma^2\mathbf{I}_L$. Thus, we can optimize $f(\boldsymbol{\gamma})$ with respect to one argument $\gamma_k, k\in \{1,2,\dots,K\}$ in one step and we iterate several times over the whole set of variables until the cost function cannot be further reduced. A random ordering is considered while optimizing to avoid dependency during detection if any. For $k\in\{1,2,\dots,K\}$, let us define $f_k^m(d) = f^m(\boldsymbol{\gamma}+d\mathbf{e}_k)$, where $\mathbf{e}_k$ is the $k^{th}$ canonical basis with a single-1 at the $k^{th}$ coordinate. 
By applying the Sherman-Morrison rank-1 update identity \cite{sherman1950adjustment} on $\mathbf{Q}_m$, we obtain 
\begin{equation}
\label{eqn:Rank1Update}
\left(\mathbf{Q}_m + d \beta_{mk}\mathbf{s}_k\mathbf{s}_k^{\text{H}} \right)^{-1}  = \mathbf{Q}_m^{-1} - d\beta_{mk}\frac{\mathbf{Q}_m^{-1}\mathbf{s}_k\mathbf{s}_k^{\text{H}}\mathbf{Q}_m^{-1}}{1+d\beta_{mk}\mathbf{s}_k^{\text{H}}\mathbf{Q}_m^{-1}\mathbf{s}_k}.
\end{equation} 

By applying the determinant identity \cite{sylvester1851xxxvii}, we can obtain
\begin{equation}
\label{eqn:DeterminantUpdate}
|\mathbf{Q}_m + d\beta_{mk}\mathbf{s}_k\mathbf{s}_k^{\text{H}}| = (1+d\beta_{mk}\mathbf{s}_k^{\text{H}}\mathbf{Q}_m^{-1}\mathbf{s}_k)|\mathbf{Q}_m|.
\end{equation}

Now we can write the overall maximum likelihood (ML) cost function in (\ref{eqn:ML_costfunc}) for each coordinate $k$ as $f_k({d}) = \sum_{m=1}^{M}f_k^m(d)$, given by
\begin{equation}
	\label{eqn:OverallCostFunction}
	\begin{aligned}
		f_k({d}) = c  + \sum_{m=1}^{M}  \bigg( 
		& \log(1+d\beta_{mk}\mathbf{s}_k^{\text{H}}\mathbf{Q}_m^{-1}\mathbf{s}_k) \\ & \ \  -d\beta_{mk}\frac{\mathbf{s}_k^{\text{H}}\mathbf{Q}_m^{-1}\mathbf{Q}_{\mathbf{Y}_m}\mathbf{Q}_m^{-1}\mathbf{s}_k}{1+d\beta_{mk}\mathbf{s}_k^{\text{H}}\mathbf{Q}_m^{-1}\mathbf{s}_k}
		\bigg),
	\end{aligned}
\end{equation}
where $c=\sum_{m=1}^{M}\left( \log|\mathbf{Q}_m| + \trace(\mathbf{Q}_m^{-1}\mathbf{Q}_{\mathbf{Y}_m})\right)$ is a constant and $\mathbf{Q}_{\mathbf{Y}_m} = \frac{\mathbf{Y}_m\mathbf{Y}_m^{\text{H}}}{N}$. Taking the derivative of $f_k(d)$ with respect to $d$ and equating to zero gives
\begin{equation}
	\label{eqn:OverallCostFunctionDerivative}
	\begin{aligned}
		f_k'({d}) = \sum_{m=1}^{M}  \bigg( 
		& \frac{\beta_{mk}\mathbf{s}_k^{\text{H}}\mathbf{Q}_m^{-1}\mathbf{s}_k}{(1+d\beta_{mk}\mathbf{s}_k^{\text{H}}\mathbf{Q}_m^{-1}\mathbf{s}_k)}  \\ & \  -d\beta_{mk}\frac{\mathbf{s}_k^{\text{H}}\mathbf{Q}_m^{-1}\mathbf{Q}_{\mathbf{Y}_m}\mathbf{Q}_m^{-1}\mathbf{s}_k}{(1+d\beta_{mk}\mathbf{s}_k^{\text{H}}\mathbf{Q}_m^{-1}\mathbf{s}_k)^2}
		\bigg) = 0,
	\end{aligned}
\end{equation}
which is a polynomial of degree $2M-1$. Hence, finding the value of $d$ which minimizes (\ref{eqn:OverallCostFunction}) requires a complexity of $\mathcal{O}(M^4L^2)$ and involves solving higher degree polynomial equations. This huge complexity calls for a low complexity design to ensure scalability of the device activity detection in cell-free massive MIMO networks. 

In the cell-free network, where the APs and devices are distributed over the cell area, the large-scale fading coefficients of the device varies significantly in magnitude between the different APs, unless the APs are equidistant from the device.  This variation can be up to the order of 50 dB in the presence of shadow fading. Exploiting these variations in channel gain in the cell-free architecture, we propose algorithms for activity detection in the next subsections. 

\subsection{Dominant AP Based Activity Detection}
\label{subsec:DominantAP}
For a device $k$, let 
\begin{equation}
m' = \underset{m}{\argmax} \{\beta_{mk} \} 
\end{equation}
be the index of the AP with which the device have the dominant large-scale fading coefficient and we call this AP the most dominant AP for the device $k$. In the proposed dominant AP based activity detection, the updates for any device is given by its corresponding dominant AP. Hence, at the CPU, we minimize the cost function with respect to the dominant AP for device $k$ and the soft information about the device $k$ from this AP is propagated to the other APs. 
The cost function of device $k$ with respect to the dominant AP $m'$ is given by
\begin{equation}
	\label{eqn:RedefinedCostFunction}
	\begin{aligned}	
		f_{k,m'}(d) \ =  \ 
		& \log(1+d\beta_{m'k}\mathbf{s}_k^{\text{H}}\mathbf{Q}_{m'}^{-1}\mathbf{s}_k) \\
		& \ \ \ -d\beta_{m'k}\frac{\mathbf{s}_k^{\text{H}}\mathbf{Q}_{m'}^{-1}\mathbf{Q}_{\mathbf{Y}_{m'}}\mathbf{Q}_{m'}^{-1}\mathbf{s}_k}{1+d\beta_{m'k}\mathbf{s}_k^{\text{H}}\mathbf{Q}_{m'}^{-1}\mathbf{s}_k}.
	\end{aligned}	
\end{equation}
Taking the derivative of (\ref{eqn:RedefinedCostFunction}) and equating it to zero, we obtain 
\begin{equation}
	d^* = \frac{\mathbf{s}^{\text{H}}_k\mathbf{Q}^{-1}_{m'}\mathbf{Q_{Y}}_{m'}\mathbf{Q}^{-1}_{m'}\mathbf{s}_k - \mathbf{s}^{\text{H}}_k\mathbf{Q}^{-1}_{m'}\mathbf{s}_k }{\beta_{m'k}(\mathbf{s}^{\text{H}}_k\mathbf{Q}^{-1}_{m'}\mathbf{s}_k)^2}.
\end{equation}

Note that $d^*$ is the minimizer of $f_{k,m'}(d)$, but need not be the minimizer of $f_{k}(d)$. To preserve the non-negativity of $\boldsymbol{\gamma}$ in (\ref{eqn:ML_func}), the optimal update step $d$ is given by $\delta=\max\{d^*,-\gamma_k\}$ and the coordinate is updated as $\gamma_k\leftarrow\gamma_k+\delta$. Using (\ref{eqn:Rank1Update}), the update step $d$ is propagated to all the sub covariance matrices $\mathbf{Q}_m, \ \forall m =1,2,\dots,M$. This procedure will be done over the whole set of random permutation of variables from the set $\{1,2,\dots,K\}$ and we iterate the entire procedure until the cost function cannot be further reduced. The proposed algorithm is summarized in Algorithm \ref{alg:AlgDomAP}. The complexity of the proposed algorithm based on dominant AP is $\mathcal{O}(IKML^2)$, where $I$ is the maximum number of iterations. The term $\mathcal{O}(L^2)$ considers the matrix-vector multiplications in Algorithm \ref{alg:AlgDomAP}. 

\floatstyle{spaceruled}
\restylefloat{algorithm}
\begin{algorithm}[!t]
	\caption{\strut Coordinate Descend Algorithm for estimating $\boldsymbol{\gamma}$}
	\begin{algorithmic}[1]
		\label{alg:AlgDomAP}
		\renewcommand{\algorithmicrequire}{\textbf{Input:}}
		\renewcommand{\algorithmicensure}{\textbf{Initialize:}}
		\REQUIRE Observations $\mathbf{Y}_m, \forall m=1,2,\dots M$, $\beta_{mk}, \forall m=1,2,\dots M , k=1,2,\dots K$
		\ENSURE  $\mathbf{Q}^{-1}_m=\sigma^{-2}\mathbf{I}_L ,  \forall m=1,2,\dots M$, $\hat{\boldsymbol{\gamma}}^0=\mathbf{0}_K$
		\STATE Compute $\mathbf{Q_{Y}}_m = \frac{1}{N}\mathbf{Y}_m\mathbf{Y}^{\text{H}}_m,  \forall m=1,2,\dots M$
		\FOR {$i = 1,2,\dots,I$ }
		\STATE Select an index set $\mathcal{K}$ from the random permutation of set $\{1,2,\dots,K\}$
		\FOR {$k\in\mathcal{K}$}
		\STATE Find the strongest link or AP for device $k$ , i.e., \\ $m' = \underset{m}{\argmax} \{\beta_{mk} \} $
		\STATE $\delta = \max \left\{ \frac{\mathbf{s}^{\text{H}}_k\mathbf{Q}^{-1}_{m'}\mathbf{Q_{Y}}_{m'}\mathbf{Q}^{-1}_{m'}\mathbf{s}_k - \mathbf{s}^{\text{H}}_k\mathbf{Q}^{-1}_{m'}\mathbf{s}_k }{\beta_{m'k}(\mathbf{s}^{\text{H}}_k\mathbf{Q}^{-1}_{m'}\mathbf{s}_k)^2},-\hat{\gamma}_k \right\} $
		\STATE$\hat{\gamma}_k^{i} = \hat{\gamma}_k^{i-1} + \delta$
		\FOR {$m=1,2,\dots,M$}
		\STATE $\mathbf{Q}^{-1}_m \leftarrow \mathbf{Q}^{-1}_m - \delta\frac{\beta_{mk}\mathbf{Q}^{-1}_m \mathbf{s}_k \mathbf{s}^{\text{H}}_k \mathbf{Q}^{-1}_m}{1+\delta\beta_{mk}\mathbf{s}^{\text{H}}_k\mathbf{Q}^{-1}_m\mathbf{s}_k}$
		\ENDFOR
		\ENDFOR
		\IF {$f(\hat{\boldsymbol{\gamma}}^i)  \geq  f(\hat{\boldsymbol{\gamma}}^{i-1}) $}
			\STATE $\hat{\boldsymbol{\gamma}}  = \hat{\boldsymbol{\gamma}}^{i-1}$  
			\STATE \textbf{break}
		\ENDIF
		\STATE $\hat{\boldsymbol{\gamma}}  = \hat{\boldsymbol{\gamma}}^{i}$  
		\ENDFOR
		\RETURN $\hat{\boldsymbol{\gamma}}$
	\end{algorithmic}
\end{algorithm}

To perform activity detection, the output from Algorithm \ref{alg:AlgDomAP} is compared against a threshold $\gamma_k^{th}$ for each device $k$ and is given by 
\begin{equation}
	\label{eqn:ActivityDetection}
	\hat{a}_k = \left\{ \begin{aligned}
		1, & \text{ if } \hat{\gamma}_k \geq \gamma_k^{th} \\
		0, & \text{ otherwise}.
	\end{aligned} \right. 
\end{equation}
Let $\hat{\mathcal{A}}=\{k \ | \ \hat{a}_k=1, \forall k\in [1,K] \ \}$ be the estimate of the set of active devices. The probability of miss detection is defined as the average of the ratio of non-detected devices and the number of active devices and the probability of false alarm is defined as the average of inactive devices declared active over inactive devices and are given respectively by
\begin{equation} \label{eqn:pfa_pmd}
	P_{md} = 1-\mathbb{E}\left\{\frac{|\mathcal{A}\cap\hat{\mathcal{A}}|_\text{c}}{|\mathcal{A}|_\text{c}}\right\} , \ \ P_{fa} = \mathbb{E}\left\{\frac{|\hat{\mathcal{A}} \setminus \mathcal{A} |_\text{c}}{K-|\mathcal{A}|_\text{c}}\right\}.
\end{equation}
The threshold $\gamma_k^{th} $ is chosen to have a desired probability of miss detection and probability of false alarm performance. 

\subsection{Clustering Based Activity Detection}
The activity detection in Algorithm~\ref{alg:AlgDomAP} uses data from one dominant AP per device and the performance improves when more antennas are used at the AP~\cite{haghighatshoar2018improved,chen2019covariance}. However, for activity detection in a cell-free network, the optimal method would be if all APs are contributing to the activity detection for all users, but this is unnecessarily computationally complex as mentioned in Sec.~\ref{sec:CoordinateDescentCostFunction}. Since only a few APs are close to each user, we consider a cluster of APs with good channels to the user. In this subsection, we consider the minimization of the cost function in~(\ref{eqn:OverallCostFunction}), by utilizing the received signals from a cluster of dominant APs for each device. Towards this, we define the function which returns the set of indices of the $T$ maximum values from the set of real numbers $\mathcal{T}$, as
$$\indmax_{\cdot,T}\{\mathcal{T}\}.$$ 
Note that the above function reduces to $\argmax$, when $T=1$.

\floatstyle{spaceruled}
\restylefloat{algorithm}
\begin{algorithm}[!t]
	\caption{\strut Clustering based coordinate descend algorithm for estimating $\boldsymbol{\gamma}$}
	\begin{algorithmic}[1]
		\label{alg:AlgCluster}
		\renewcommand{\algorithmicrequire}{\textbf{Input:}}
		\renewcommand{\algorithmicensure}{\textbf{Initialize:}}
		\REQUIRE Observations $\mathbf{Y}_m, \forall m=1,2,\dots M$, $\beta_{mk}, \forall m=1,2,\dots M , k=1,2,\dots,K$
		\ENSURE  $\mathbf{Q}^{-1}_m=\sigma^{-2}\mathbf{I}_L ,  \forall m=1,2,\dots,M$, $\hat{\boldsymbol{\gamma}}^0=\mathbf{0}_K$ \\
		\STATE Compute $\mathbf{Q_{Y}}_m = \frac{1}{N}\mathbf{Y}_m\mathbf{Y}^{\text{H}}_m,  \forall m=1,2,\dots,M$ \\
		Compute $\mathcal{M}_k = \indmax_{m,T} \ \{ \beta_{mk}\}  \forall  k=1,2,\dots,K$
		\FOR {$i = 1,2,\dots,I$ }
			\STATE Select an index set $\mathcal{K}$ from the random permutation of set $\{1,2,\dots,K\}$
			\FOR {$k\in\mathcal{K}$}
				\FOR {$m\in\mathcal{M}_k$ }
					\STATE Compute $ a_m = \beta_{mk}\mathbf{s}_k^{\text{H}}\mathbf{Q}_m^{-1}\mathbf{s}_k $ and \\
					$b_m = \beta_{mk}\mathbf{s}_k^{\text{H}}\mathbf{Q}_m^{-1}\mathbf{Q}_{\mathbf{Y}_m}\mathbf{Q}_m^{-1}\mathbf{s}_k$
				\ENDFOR
				
				\STATE \label{alg:PolyEqn1} Solve the polynomial equation \\ $ \begin{aligned}
				f'_{k,T}(d) = & \sum_{m\in\mathcal{M}_k} \bigg(((a_m+b_m) +a_m^2d).  \\ & \prod_{m'\in\mathcal{M}_k\setminus\{m\}} (1 + 2a_{m'}d + a_{m'}^2d^2) \bigg) = 0
				\end{aligned}   $
				
				\STATE Compute  $\mathcal{D} = \{d : \ f'_{k,T}(d) = 0, \ \Im(d)=0, \ \Re(d)\geq-\gamma_k \}\cup\{-\gamma_k\}$
		
				\STATE Let $f_{k,T}(d) = \sum_{m\in\mathcal{M}_k} \left(\log(1+da_m) - \frac{db_m}{1+da_m}\right)$. \\
				Compute $\delta = \argmin_{d\in\mathcal{D}} f_{k,T}(d)$.
		
				\STATE$\hat{\gamma}_k^i = \hat{\gamma}_k^{i-1} + \delta$
				\FOR {$m=1,2,\dots,M$}
					\STATE $\mathbf{Q}^{-1}_m \leftarrow \mathbf{Q}^{-1}_m - \delta\frac{\beta_{mk}\mathbf{Q}^{-1}_m \mathbf{s}_k \mathbf{s}^{\text{H}}_k \mathbf{Q}^{-1}_m}{1+\delta\beta_{mk}\mathbf{s}^{\text{H}}_k\mathbf{Q}^{-1}_m\mathbf{s}_k}$
				\ENDFOR
			\ENDFOR
			\IF {$f(\hat{\boldsymbol{\gamma}}^i)  \geq  f(\hat{\boldsymbol{\gamma}}^{i-1}) $}
				\STATE $\hat{\boldsymbol{\gamma}}  = \hat{\boldsymbol{\gamma}}^{i-1}$  
				\STATE \textbf{break}
			\ENDIF
			\STATE $\hat{\boldsymbol{\gamma}}  = \hat{\boldsymbol{\gamma}}^{i}$  
		\ENDFOR
		\RETURN $\hat{\boldsymbol{\gamma}}$
	\end{algorithmic}
\end{algorithm}

Let
\begin{equation}
	\mathcal{M}_k = \indmax_{m,T} \ \{ \beta_{mk}\}, 
\end{equation}
be the cluster of $T<M$ dominant APs of the device $k$. For $m\in\mathcal{M}_k$, define
\begin{align}
a_m & = \beta_{mk}\mathbf{s}_k^{\text{H}}\mathbf{Q}_m^{-1}\mathbf{s}_k \\
b_m & = \beta_{mk}\mathbf{s}_k^{\text{H}}\mathbf{Q}_m^{-1}\mathbf{Q}_{\mathbf{Y}_m}\mathbf{Q}_m^{-1}\mathbf{s}_k.
\end{align}

To minimize the cost function (\ref{eqn:OverallCostFunction}) by utilizing the signals from the $T$ dominant APs for the user $k$, we redefine the cost function as 

\begin{equation}
\label{eqn:NewCostFunc}	
f_{k,T}(d) = \sum_{m\in\mathcal{M}_k} \left(\log(1+da_m) - \frac{db_m}{1+da_m}\right).
\end{equation}
Taking the derivative of (\ref{eqn:NewCostFunc}) with respect to $d$, yields 
\begin{equation}
	\label{eqn:DiffCostFunc}
	f_{k,T}'({d}) = \sum_{m\in\mathcal{M}_k} \frac{a_m}{1+da_m} + \frac{b_m}{(1+da_m)^2}. 
\end{equation}
Equating (\ref{eqn:DiffCostFunc}) to zero yields
\begin{equation}
\label{eqn:polyEquation}	
\begin{aligned}
\sum_{m\in\mathcal{M}_k} \bigg(  & ((a_m+b_m) +a_m^2d)  \\
& \prod_{m'\in\mathcal{M}_k\setminus\{m\}} (1 + 2a_{m'}d + a_{m'}^2d^2)\bigg) = 0 
\end{aligned}
\end{equation}
which is a polynomial equation in $d$ of degree $2T-1$. Let 
\begin{equation}
	\mathcal{D} = \{d:f'_{k,T}(d) = 0, \Im(d)=0,\Re(d)\geq-\gamma_k \}\cup\{-\gamma_k\},
\end{equation}
be the set of real roots of (\ref{eqn:polyEquation}) and compute
\begin{equation}
\delta = \argmin_{d\in\mathcal{D}} f_{k,T}(d). 
\end{equation}
The value $-\gamma_k$ is added to the set $\mathcal{D}$ to preserve the positivity of  $\boldsymbol{\gamma}$ in (\ref{eqn:ML_func}) and the coordinate is updated as $\gamma_k\leftarrow\gamma_k+\delta$. The updating of sub-covariance blocks is carried out as explained in Sec. \ref{subsec:DominantAP}. The proposed algorithm is outlined in Algorithm \ref{alg:AlgCluster} and the activity detection can be performed using (\ref{eqn:ActivityDetection}).

When $T=1$, the clustering based algorithm reduces to Algorithm~\ref{alg:AlgDomAP}. For $T=2$, we have degree 3 polynomial equation in (\ref{eqn:polyEquation}) and the roots can be solved in closed form~\cite{holmes2002use}. For $T>2$, we have polynomials of degree 5 and higher and there exists no closed form solutions for the roots~\cite{alekseev2004abel}. For $T>2$, the approximate roots of the polynomial in (\ref{eqn:polyEquation}) can be found by finding the eigen values of the companion matrix formed using the coefficients \cite[Ch. 6]{mcnamee2007numerical} and the computation complexity is $\mathcal{O}(T^3)$ \cite{aberth1973iteration}. Thus the overall complexity of the Algorithm \ref{alg:AlgCluster} is $\mathcal{O}(IK(TL^2+T^3+ML^2))$. The term $T^3$ corresponds to the complexity for finding the coefficients of (\ref{eqn:polyEquation}), which can be computed using $2T$ point convolution. The term $TL^2$ and $ML^2$ corresponds to the complexities associated with computation of coefficients $a_m,~b_m$ and updating of covariance matrices, respectively. By introducing the clustering, we can identify a suitable tradeoff between performance (large $ T $) and complexity (small $ T $).

\subsection{Parallel Architecture of Algorithms}
\label{sec:LessComplexAlgorithms}
In Algorithms \ref{alg:AlgDomAP} and \ref{alg:AlgCluster}, the update of each user is done sequentially irrespective of whether the user is active or not. In mMTC applications, the probability, $\epsilon$, of device being active is very small, and thus on an average the number of active devices being active at any time is $K\epsilon$, which is much smaller. Hence, the sub-covariance matrices in Algorithm \ref{alg:AlgCluster} do not change much while updating for an inactive user.
Let $\mathcal{K}=$ random permutation of set $\{1,2,\dots,K\}$. Divide $\mathcal{K}$ into $G$ random disjoint groups i.e., $\mathcal{K}=\mathcal{K}_1\cup\mathcal{K}_2\cup\cdots\cup\mathcal{K}_G$. For each group $\mathcal{K}_g$, compute the coefficients $a_m,~b_m$ once and find the update $\delta$. Once the updates for each user in group $g$ are obtained, update the covariance matrix and continue to update for next group. This operation can be done in parallel and hence such a parallel architecture can save up to $G$ times the time required for Algorithm \ref{alg:AlgCluster}. The proposed algorithm is summarized in Algorithm \ref{alg:AlgClusParallel}. 

\floatstyle{spaceruled}
\restylefloat{algorithm}
\begin{algorithm}[!t]
	\caption{\strut Clustering based coordinate descend parallel architecture algorithm for estimating $\boldsymbol{\gamma}$}
	\begin{algorithmic}[1]
		\label{alg:AlgClusParallel}
		\renewcommand{\algorithmicrequire}{\textbf{Input:}}
		\renewcommand{\algorithmicensure}{\textbf{Initialize:}}
		\REQUIRE Observations $\mathbf{Y}_m, \forall m=1,2,\dots M$, $\beta_{mk}, \forall m=1,2,\dots M , k=1,2,\dots,K$
		\ENSURE  $\mathbf{Q}^{-1}_m=\sigma^{-2}\mathbf{I}_L ,  \forall m=1,2,\dots,M$, $\hat{\boldsymbol{\gamma}}^0=\mathbf{0}_K$ \\
		\STATE Compute $\mathbf{Q_{Y}}_m = \frac{1}{N}\mathbf{Y}_m\mathbf{Y}^{\text{H}}_m,  \forall m=1,2,\dots,M$ \\
		Compute $\mathcal{M}_k = \indmax_{m,T} \ \{ \beta_{mk}\}  \forall  k=1,2,\dots,K$
		\FOR {$i = 1,2,\dots,I$ }
			\STATE Select an index set $\mathcal{K}$ from the random permutation of set $\{1,2,\dots,K\}$ and find random disjoint sets such that $\mathcal{K} = \mathcal{K}_1\cup\mathcal{K}_2\cup\dots\cup\mathcal{K}_G$
			\FOR {$g=1,2,\dots,G$} 
				\FOR {$m\in\mathcal{M}_k$ and $k\in\mathcal{K}_g$ }
					\STATE Compute $ a_{mk} = \beta_{mk}\mathbf{s}_k^{\text{H}}\mathbf{Q}_m^{-1}\mathbf{s}_k $ and \\
					$b_{mk} = \beta_{mk}\mathbf{s}_k^{\text{H}}\mathbf{Q}_m^{-1}\mathbf{Q}_{\mathbf{Y}_m}\mathbf{Q}_m^{-1}\mathbf{s}_k$
				\ENDFOR
				
				\FOR {$k\in\mathcal{K}_g$  \fbox{Parallel processing} } 
					\STATE \label{alg:PolyEqn2} Solve the polynomial equation \\ $ \begin{aligned}
					f'_{k,T}(d) = & \sum_{m\in\mathcal{M}_k} \bigg(((a_m+b_m) +a_m^2d).  \\ & 	\prod_{m'\in\mathcal{M}_k\setminus\{m\}} (1 + 2a_md + a_m^2d^2) \bigg) = 0
					\end{aligned}   $
					\STATE Compute  $\mathcal{D} = \{d : \ f'_{k,T}(d) = 0, \ \Im(d)=0, \ \Re(d)\geq-\gamma_k \}\cup\{-\gamma_k\}$
					\STATE Compute $\delta_k = \argmin_{d\in\mathcal{D}} f_{k,T}(d)$, for \\
					$f_{k,T}(d) = \sum_{m\in\mathcal{M}_k} \left(\log(1+da_{mk}) - \frac{db_{mk}}{1+da_{mk}}\right)$.
					\STATE$\hat{\gamma}_k^i = \hat{\gamma}_k^{i-1} + \delta_k$
				\ENDFOR
				\FOR {$m=1,2,\dots,M$ and $k\in\mathcal{K}_g$}
					\STATE $\mathbf{Q}^{-1}_m \leftarrow \mathbf{Q}^{-1}_m - \delta_k\frac{\beta_{mk}\mathbf{Q}^{-1}_m \mathbf{s}_k \mathbf{s}^{\text{H}}_k \mathbf{Q}^{-1}_m}{1+\delta_k\beta_{mk}\mathbf{s}^{\text{H}}_k\mathbf{Q}^{-1}_m\mathbf{s}_k}$
				\ENDFOR
			\ENDFOR
			
			\IF {$f(\hat{\boldsymbol{\gamma}}^i)  \geq  f(\hat{\boldsymbol{\gamma}}^{i-1}) $}
				\STATE $\hat{\boldsymbol{\gamma}}  = \hat{\boldsymbol{\gamma}}^{i-1}$  
				\STATE \textbf{break}
			\ENDIF
			
			\STATE $\hat{\boldsymbol{\gamma}}  = \hat{\boldsymbol{\gamma}}^{i}$  
		\ENDFOR
		\RETURN $\hat{\boldsymbol{\gamma}}$
	\end{algorithmic}
\end{algorithm}

\subsection{Convergence of the Algorithms}
\label{sec:convergence}
The cost function $ f(\boldsymbol{\gamma}) $ in (\ref{eqn:ML_costfunc}) is a sum of concave ($\log$) and convex (trace) functions in $\boldsymbol{\gamma}$. Hence, closed-form expression to find the global minimum of (\ref{eqn:ML_costfunc}) does not exist and in this paper, we use sub-optimal algorithms to find a local minimum of (\ref{eqn:ML_costfunc}). First, we look at each term in $ f(\boldsymbol{\gamma}) $. $ \trace\left(\mathbf{Q}_m^{-1}\frac{\mathbf{Y}_m\mathbf{Y}_m^{\text{H}}}{N}\right), \ \forall m=1,2,\cdots,M $ are convex and hence, are bounded from below. Also from (\ref{eqn:RankOneUpdates}), we can see that $ \mathbf{Q}_m(\boldsymbol{\gamma}) $ is a sum of positive semi-definite rank-one updates to a positive definite matrix $ \sigma^2\mathbf{I}_L $. Hence $ \lvert \mathbf{Q}_m(\boldsymbol{\gamma}) \rvert > 0, \ \forall m=1,2,\cdots,M $. Thus the cost function  $ f(\boldsymbol{\gamma}) $ is bounded from below. 
With each iteration in the proposed algorithms, the cost function $ f(\boldsymbol{\gamma}) $ is non-increasing. Hence, the proposed algorithms, where the cost function $ f(\boldsymbol{\gamma}) $ is monotonically decreasing and being bounded from below, is guaranteed to converge to a local minimum.

\section{Simulation Results}
\label{sec:SimulationResults}
In this section, we characterize the massive connectivity in cell-free massive MIMO architectures and plot the performance of massive activity detection in cell-free massive MIMO with our proposed algorithms. We consider the receiver operating characteristic (ROC) characterized by probability of miss detection and probability of false alarm, as the performance measures for activity detection. 

\subsection{Simulation Model}
We consider a square area wrapped around the edges to mimic a network with infinite area and to avoid boundary effects. The $M$ APs are arbitrarily and independently distributed. The $K$ users are considered to be uniformly distributed in the network. We will compare grant-free random access in cell-free and co-located architectures and hence, for co-located case, we assume the AP is at the center of the network. We consider such a simulation area with $K=400$ devices, the activation probability $\epsilon=0.1$ and the signature sequence length $L=40$. The following micro-cell propagation model used in \cite{bjornson2019making} is considered for large-scale fading coefficient $\beta_{mk}$:
\begin{equation}
\label{eqn:PathLoss}
\beta_{mk}[\text{dB}] = -30.5 - 36.7 \log_{10} \left( \frac{d_{mk}}{1\text{m}}\right) + F_{mk}
\end{equation}
where $F_{mk}\sim\mathcal{N}(0,\sigma_{sh}^2)$ is the shadow fading component with variance $\sigma_{sh}^2$, and $d_{mk}$ is the horizontal distance between the $k^{th}$ user and the $m^{th}$ AP in meters ignoring their height differences. We use the same propagation model for the co-located massive MIMO case to ensure that the performance differences are caused by differences in technology characteristics instead of propagation model differences. Note that the shadow fading effects can be larger in co-located architecture. The maximum transmit power for a device is 200 mW and noise power $\sigma^2=-109 \ \text{dBm}$. We consider a coherence block of 1 ms and 200 kHz, such that $ \tau_c=200 $ symbols can be transmitted. However, due to large number of users in the system $ K=400>\tau_c $, we assign non-orthogonal pilot sequence to each user. We reserve 20\% of available symbols for pilot sequences. Hence in this paper, we consider $ L=40 $ as the signature sequence length. The signature sequence of each device is assumed to be drawn from the Gaussian distribution, $ \mathcal{CN}(\mathbf{0},\mathbf{I}_L)$. 

\subsection{Results}
\label{sec:results}

\begin{figure}[t]
	\centering
	\includegraphics[scale=0.45]{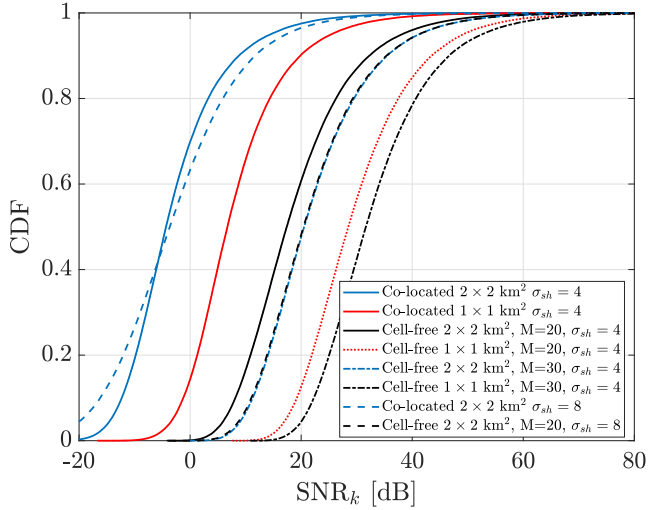}
	\caption{Active device SNR}
	\label{fig:ActiveDeviceSNR}
\end{figure}

First, we look at the SNR achieved at the AP (dominant AP for cell-free) by an active device $k$ transmitting at a power of 200 mW for co-located and cell-free massive MIMO networks of $2\times2 \  \text{km}^2$ and $1\times1 \ \text{km}^2$ cell area sizes and is shown in Fig. \ref{fig:ActiveDeviceSNR} for different scenarios. The plot shows that there is a significant gap in the SNR achieved for co-located and cell-free massive MIMO and hence the outage probability (probability of not achieving a certain SNR target) is less in the cell-free massive MIMO scenario \cite{demir2021foundations}, thereby enabling services to many devices. Due to the distributed topology of APs in the cell-free  scenario, a device is highly likely to be close to one of the APs, thus providing a stronger channel gain which improves the SNR. This gain in SNR is referred to as macro-diversity gain of cell-free systems~\cite[pp. 331]{demir2021foundations}. Also, in cell-free networks when more APs are deployed in the cell area, the SNR achieved at the dominant AP increases. Fig. \ref{fig:ActiveDeviceSNR} shows that the co-located MIMO network is sensitive to shadow fading while the cell-free network is robust to shadow fading. Co-located network is likely to have larger shadow fading parameter compared to cell-free network according to standard 3GPP channel models \cite{ngo2017cell,bjornson2019making}.

Next, we look at the performance of the proposed algorithms for activity detection in cell-free massive MIMO network. For simulations, we consider an SNR target at the dominant AP such that 95\% of the active devices will be able to achieve the desired SNR and hence access the network. Referring to Fig.~\ref{fig:ActiveDeviceSNR}, target SNR can be computed by finding the SNR where CDF=$ 0.05 $ for each test cases. The variance of shadow fading parameter is $\sigma_{sh}^2=4$. The ROC curve is plotted for different thresholds. We have considered $I=10$ as the maximum number of iterations and $10^5$ Monte-Carlo trials.

\begin{figure}[t]
	\centering
	\includegraphics[scale=0.45]{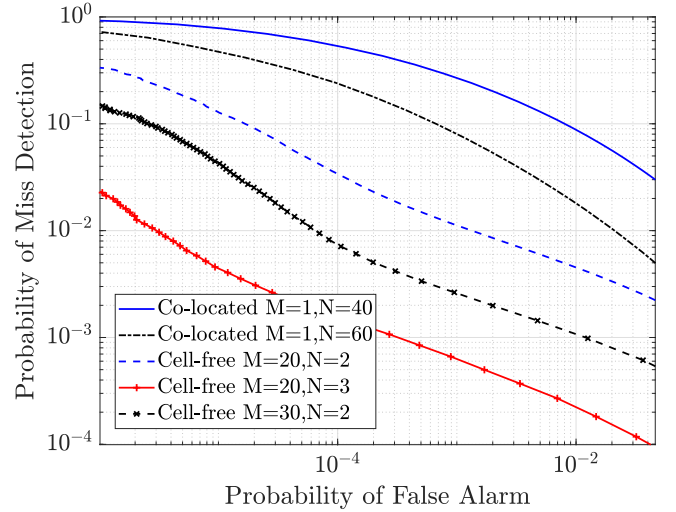}
	\caption{Performance in $2\times2 \ \text{km}^2$ cell area. $K=400$ users, activation probability $\epsilon=0.1$, sequence length $L=40$.}
	\label{fig:Performance_2sqkm}
\end{figure}

\begin{figure}[t]
	\centering
	\includegraphics[scale=0.45]{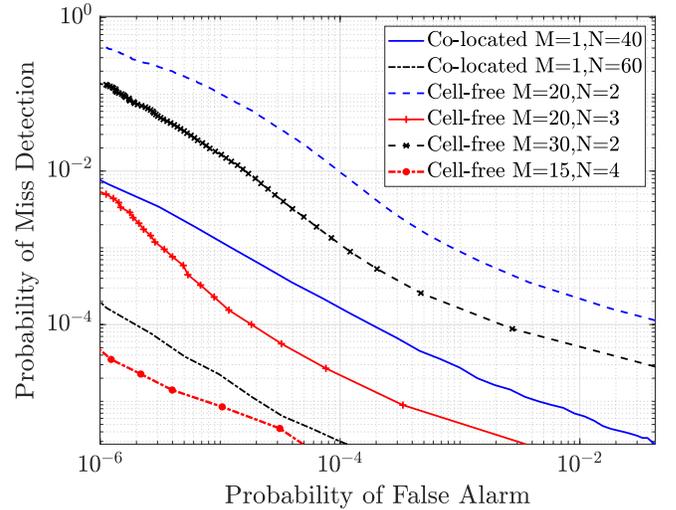}
	\caption{Same as Fig. \ref{fig:Performance_2sqkm} but for $1\times1 \ \text{km}^2$ cell area.}
	\label{fig:Performance_1sqkm}
\end{figure}

The performance of Algorithm \ref{alg:AlgDomAP} is given in Fig. \ref{fig:Performance_2sqkm} and Fig. \ref{fig:Performance_1sqkm} for $2\times2 \  \text{km}^2$ and $1\times1 \ \text{km}^2$ cell area sizes, respectively. From the plots, it can be seen that, when the number of antennas per AP, $ N $ is increased, the activity detection performance improves due to the improvement in the spatial resolution of the devices. For low-power applications like mMTC, co-located MIMO is highly sensitive to the receive SNR and the performance degrades significantly with an increase in cell area. From the plots, it can be seen that the device activity detection performance is better in cell-free massive MIMO networks compared to co-located MIMO networks, as the cell-area increases. Hence, cell-free network can provide a broad coverage in mMTC applications. Note that for co-located architecture, the network will have higher shadow fading effects compared to cell-free architecture which will further reduce the performance owing to the reduction in the SNR.
When the AP density in the network defined by $ \frac{M}{\text{cell area}} $ increases, the SNR at the dominant AP increases, thereby improving the activity detection performance as shown in Figs. \ref{fig:Performance_2sqkm} and \ref{fig:Performance_1sqkm}.

\begin{figure}[t]
	\centering
	\includegraphics[scale=0.45]{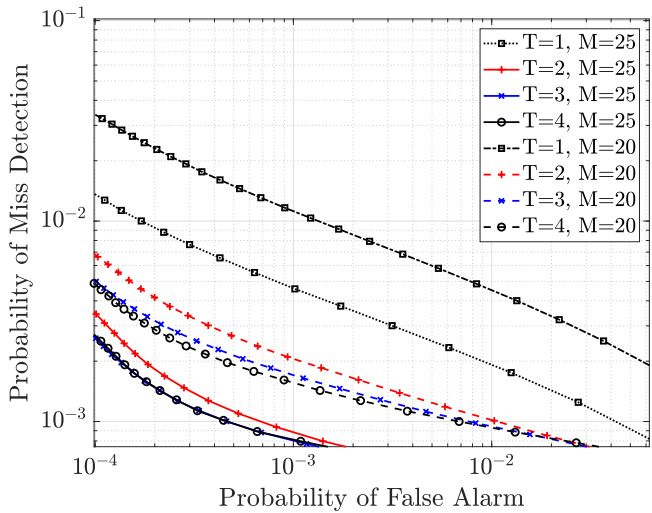}
	\caption{Activity detection performance with $ 2\times2 \ \text{km}^2$ cell area, $\ N=2, \ K=400, \ L=40, \ \epsilon=0.1 $.}
	\label{fig:Performance_2KM}
\end{figure}

The performance of the Algorithm \ref{alg:AlgCluster} for activity detection in cell-free massive MIMO network is given in Fig. \ref{fig:Performance_2KM} for $2\times2 \ \text{km}^2$ cell area size. From the plots, we can see that when we consider information from multiple dominant APs ($ T>1 $) for activity detection, the performance improves compared to considering only the most dominant AP. Moreover, to study the performance gain from the dominant APs and to have a fair comparison we consider two cases, one with $ M=20 $ and other with $ M=25 $, such that the AP density is higher in the latter case. From the plot, it can be seen that the performance gain for activity detection improves with the dominance of the APs. However, considering information from three or more dominant APs, the performance is not improving by a significant amount compared to considering information from the two dominant APs. Also, with $T=2$, the algorithm is computationally efficient as closed form expressions are available for solving a degree 3 polynomial equation.

\begin{figure}[t]
	\centering
	\includegraphics[scale=0.45]{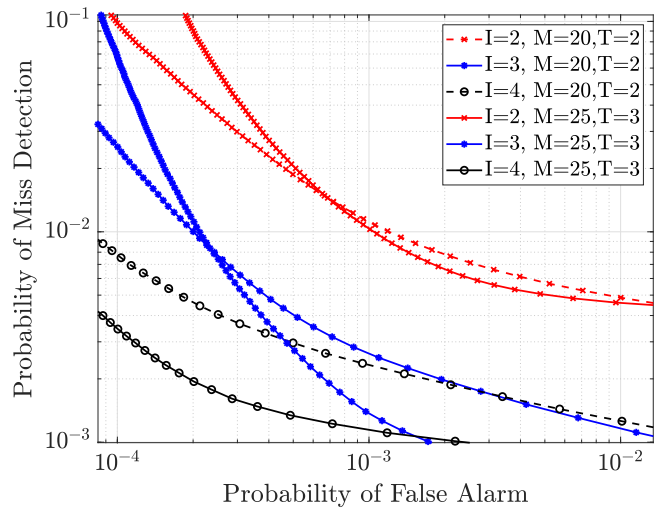}
	\caption{Convergence of algorithm  \ref{alg:AlgCluster} with $ 2\times2 \ \text{km}^2$ cell area, $\ N=2, \ K=400, \ L=40$.}
	\label{fig:Performance_2sqkm_Convergence}
\end{figure}

The convergence of the algorithm with the number of iterations $I$ is plotted in Fig. \ref{fig:Performance_2sqkm_Convergence} for different scenarios. Since the cost function (\ref{eqn:ML_costfunc}) is non-increasing and bounded from below, when the proposed algorithms are used, the activity detection algorithms are able to achieve convergence (to a local minimum).

\begin{figure}[t]
	\centering
	\includegraphics[scale=0.45]{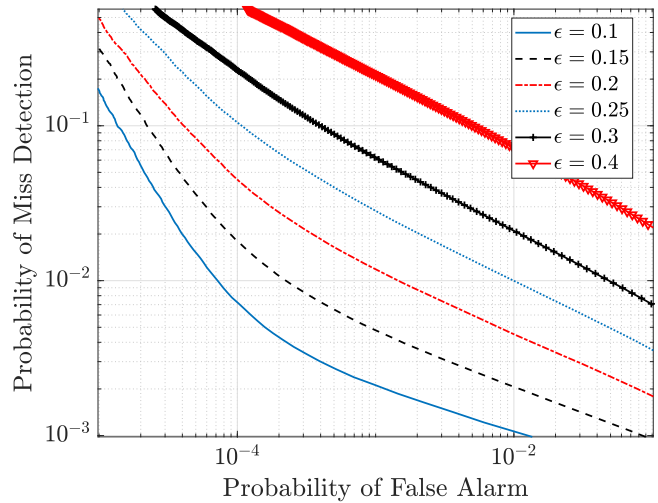}
	\caption{Activity detection performance with $ 2\times2 \ \text{km}^2$ cell area, $ M=20, \ N=2, \ K=400, \ L=40, \ T=2 $ for different device activation probability $\epsilon$. }
	\label{fig:Performance_WithDiffEps}
\end{figure}

In Fig. \ref{fig:Performance_WithDiffEps}, we study the impact of different activation probabilities on the activity detection performance of Algorithm \ref{alg:AlgCluster}. From the plot, it can be seen that, we can still achieve a nominal performance for activity detection when $ K\epsilon> L $, while using the AMP-based approach, the performance degrades severely in such situations \cite{senel2017device,donoho2009message}. Hence, the proposed algorithms can support activity detection when total number of active devices are greater than the signature sequence length.

\begin{figure}[t]
	\centering
	\includegraphics[scale=0.45]{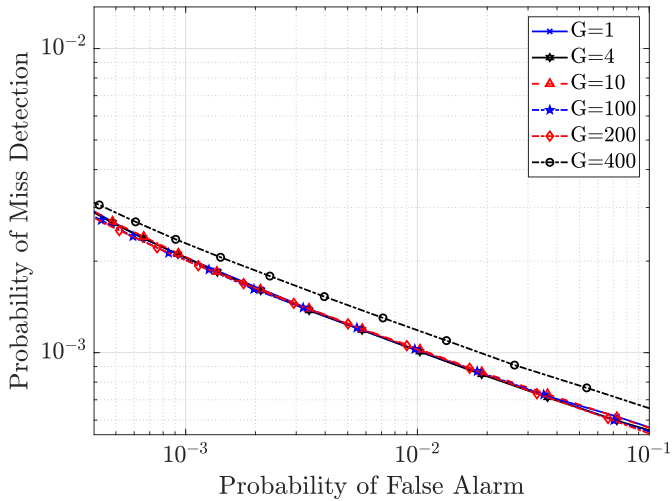}
	\caption{Activity detection performance with Algorithm \ref{alg:AlgClusParallel}  for $ 2\times2 \ \text{km}^2$ cell area, $ M=20, \ N=2, \ K=400, \ L=40, \ \epsilon=0.1, \ T=2 $.}
	\label{fig:Algorithm3_Performance}
\end{figure}

The performance of Algorithm \ref{alg:AlgClusParallel} is shown in Fig. \ref{fig:Algorithm3_Performance}. The superiority of Algorithm \ref{alg:AlgClusParallel} is in the parallelism and hence, the computational time can be cut by $ G $ times. From the plot, it can be seen that the performance of Algorithm-3 is similar to Algorithm-2 due to the fact that we have more inactive users compared to active users and hence the cost function will not change much while updating the sub-covariance matrices of inactive users. However, when we apply complete parallelism $ G=K $, we slightly loose in performance as shown in the Fig.~\ref{fig:Algorithm3_Performance}.  

\begin{figure}[t]
	\centering
	\includegraphics[scale=0.45]{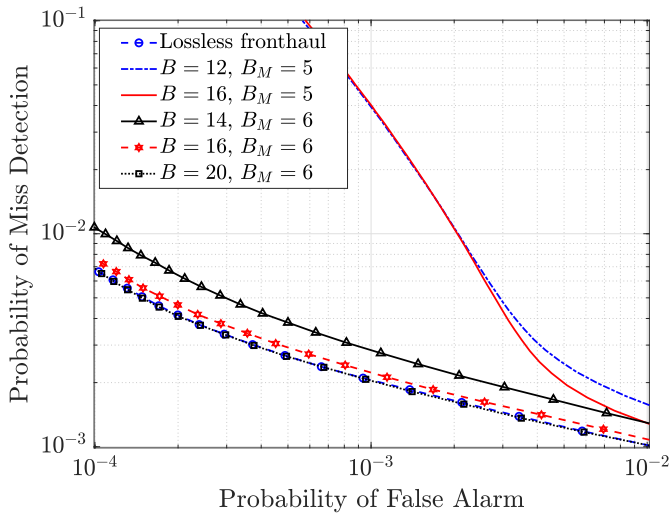}
	\caption{Activity detection performance with $ 2\times2 \ \text{km}^2$ cell area, $ M=20, \ N=2, \ K=400, \ L=40, \ T=2, \ \epsilon=0.1 $ with a capacity limited fronthaul.}
	\label{fig:lossy_fronthaul}
\end{figure}

\subsection{Capacity Limited Fronthauls}
\label{sec:lossy_fronthaul}
In practice, the fronthauls connecting the APs to the CPU are capacity limited and causes degradation in the performance. The data from APs will be quantized before sending to the CPU. We consider each complex value to be send from each AP is represented using $ B $ bits in floating point representation. Out of $\frac{B}{2}$ bits available for a real value, $ B_M $ are assigned for mantissa representation and rest for the exponent representation. The performance of a capacity limited fronthaul, for Algorithm \ref{alg:AlgCluster} is plotted in Fig.~\ref{fig:lossy_fronthaul}. The activity detection performance improves as the number of bits $ B $ increases. From the plot, when Algorithm \ref{alg:AlgCluster} is used with $ T=2 $, to achieve the optimal performance of a lossless fronthaul, we require $ B=20 $ bits per complex value. The contemporary fronthaul technology has a capacity of 10 Gbps or more, so even if the capacity is not infinite, it won't be the limiting factor for the operations that we are considering.

\section{Conclusion}
\label{sec:Conclusion}
In this paper, we analyzed the grant-free random access scenario in cell-free massive MIMO networks. The paper formulates a novel activity detection problem and proposed algorithms for activity detection based on clustering approach. We show that for low-powered applications like mMTC, co-located massive MIMO is highly sensitive to the large SNR variations. The cell-free massive MIMO network is robust against the shadow fading effects providing macro diversity gains and, hence, can provide better coverage for mMTC applications. Simulation results show that the macro diversity gain offered by cell-free network improves the activity detection performance.


\bibliographystyle{IEEEtran}
\bibliography{references}

\vfill

\begin{IEEEbiography}[{\includegraphics[width=1in,height=1.25in,clip,keepaspectratio]{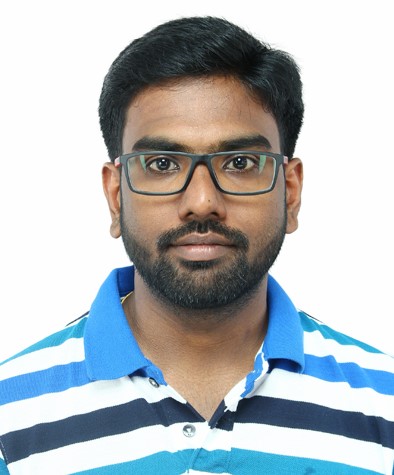}}]{Unnikrishnan Kunnath Ganesan}
(Graduate Student Member, IEEE) received the B.Tech. degree in electronics and communication engineering from the University of Calicut, Kerala, India, in 2011, and the M.E. degree in telecommunication engineering from Indian Institute of Science, Bengaluru, India, in 2014. From 2014 to 2017, he worked as a Modem Systems Engineer with Qualcomm India Private Ltd., Bengaluru. From 2017 to 2019, he worked
as a Senior Firmware Engineer with Intel. He is currently pursuing the Ph.D. degree with the Department of Electrical Engineering (ISY), Linköping University, Sweden. His primary research interests include MIMO wireless communications, space-time coding, network coding, and information theory.
\end{IEEEbiography}

\vfill

\begin{IEEEbiography}[{\includegraphics[width=1.2in,height=1.25in,clip,keepaspectratio]{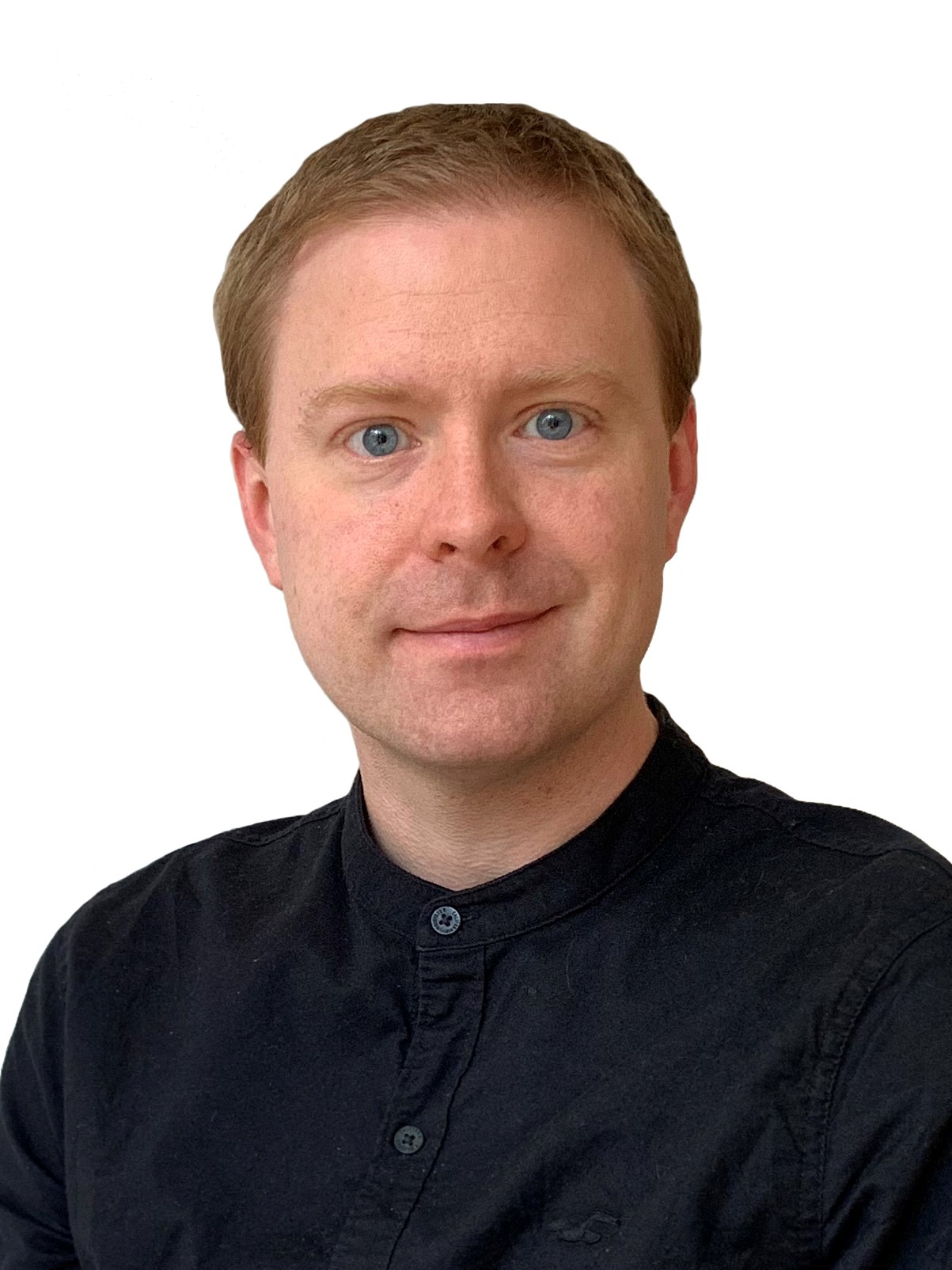}}]{Emil Bj\"ornson}
(Senior Member, IEEE) received the M.S. degree in engineering mathematics from Lund University, Sweden, in 2007, and the Ph.D. degree
in telecommunications from the KTH Royal Institute of Technology, Sweden, in 2011. From 2012 to 2014, he held a joint post-doctoral
position at Alcatel-Lucent Chair on Flexible Radio, Supelec, France, and the KTH Royal Institute of Technology. He joined Linköping University, Sweden, in 2014, where he is currently an Associate Professor. In September 2020, he became a parttime Visiting Full Professor at the KTH Royal Institute of Technology. He has authored the textbooks Optimal Resource Allocation in Coordinated
Multi-Cell Systems in 2013, Massive MIMO Networks: Spectral, Energy, and Hardware Efficiency in 2017, and Foundations of User-Centric CellFree Massive MIMO in 2021. He is dedicated to reproducible research and has made a large amount of simulation code publicly available. He has performed MIMO research for over 15 years, his articles have received more than 14 000 citations, and has filed more than 20 patent applications. He is a host of the podcast wireless future and has a popular YouTube channel. He performs research on MIMO communications, radio resource allocation, machine learning for communications, and energy efficiency.

Dr. Björnson has been a member of Online Editorial Team of IEEE TRANSACTIONS ON WIRELESS COMMUNICATIONS since 2020. He has received the
2014 Outstanding Young Researcher Award from IEEE ComSoc EMEA, the 2015 Ingvar Carlsson Award, the 2016 Best Ph.D. Award from EURASIP,
the 2018 IEEE Marconi Prize Paper Award in Wireless Communications, the 2019 EURASIP Early Career Award, the 2019 IEEE Communications
Society Fred W. Ellersick Prize, the 2019 IEEE Signal Processing Magazine Best Column Award, the 2020 Pierre-Simon Laplace Early Career Technical Achievement Award, the 2020 CTTC Early Achievement Award, and the 2021 IEEE ComSoc RCC Early Achievement Award. He also coauthored articles that received Best Paper Awards at the conferences, including WCSP 2009, IEEE CAMSAP 2011, IEEE SAM 2014, IEEE WCNC 2014, IEEE ICC 2015, and WCSP 2017. He has been on the Editorial Board of IEEE TRANSACTIONS ON COMMUNICATIONS since 2017. He has been an Area Editor in IEEE Signal Processing Magazine since 2021. He has also been a guest editor of multiple special issues.
\end{IEEEbiography}

\newpage

\begin{IEEEbiography}[{\includegraphics[width=1.1in,height=1.35in,clip,keepaspectratio]{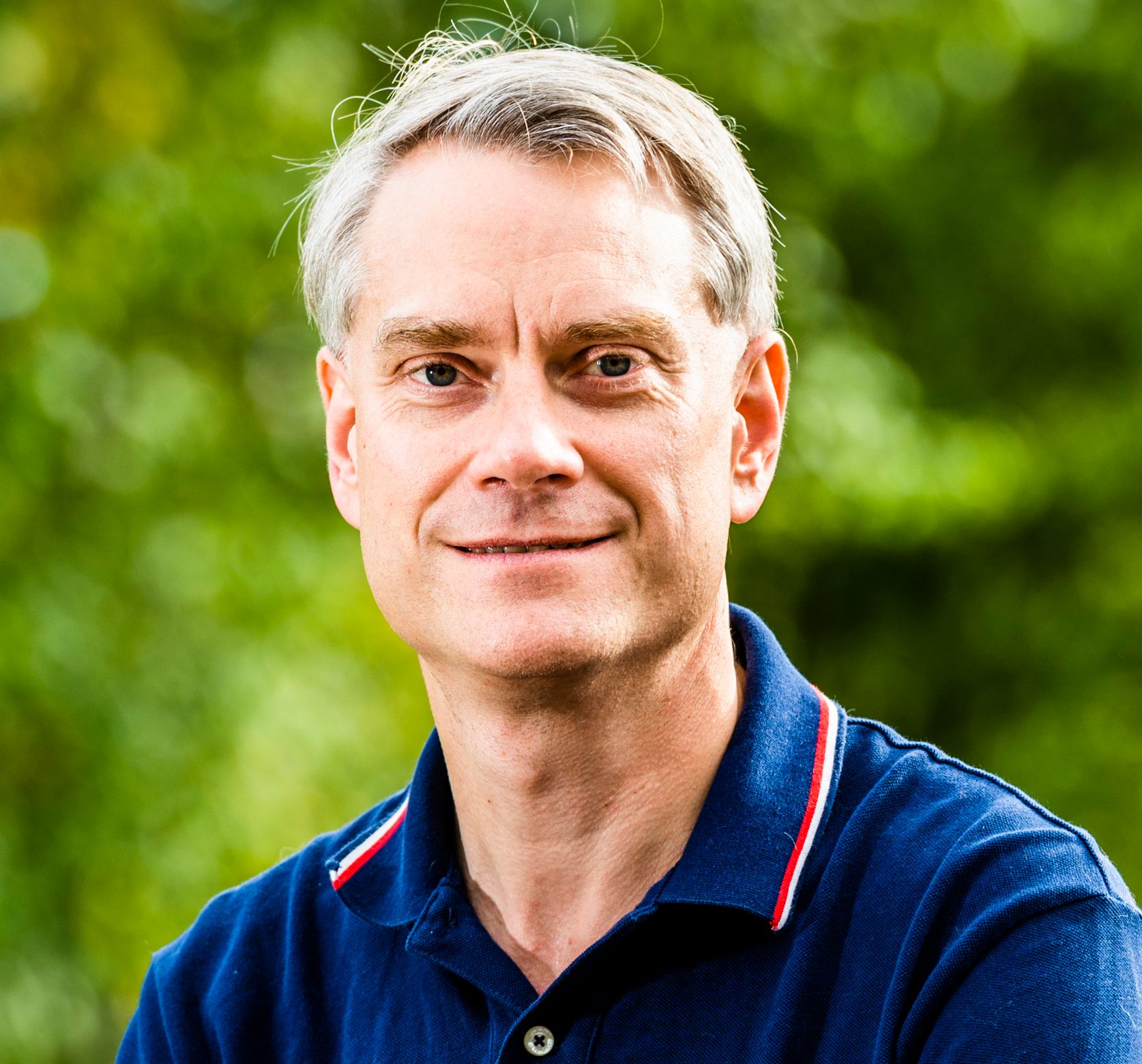}}]{Erik G. Larsson}
(Fellow, IEEE) received the Ph.D. degree from Uppsala University, Uppsala, Sweden, in 2002. He is currently a Professor of communication systems with Linköping University (LiU), Linköping, Sweden. He was with the KTH Royal Institute of Technology, Stockholm, Sweden; The
George Washington University, USA; the University of Florida, USA; and Ericsson Research, Sweden. His main professional interests are within the areas of wireless communications and signal processing. He has coauthored Space-Time Block Coding for Wireless Communications (Cambridge University Press, 2003) and Fundamentals of Massive MIMO (Cambridge University Press, 2016).

He is currently a member of the IEEE TRANSACTIONS ON WIRELESS COMMUNICATIONS Steering Committee and an Editorial Board Member of IEEE Signal Processing Magazine. He served as the Chair for the IEEE Signal Processing Society SPCOM Technical Committee (2015–2016), the Chair for the IEEE WIRELESS COMMUNICATIONS LETTERS Steering Committee (2014–2015), the General and Technical Chair for the Asilomar SSC Conference (2015 and 2012), the Technical Co-Chair for the IEEE Communication Theory Workshop (2019), and a member for the IEEE Signal Processing Society Awards Board (2017–2019). He was an Associate Editor for, among others, IEEE TRANSACTIONS ON COMMUNICATIONS (2010–2014) and IEEE TRANSACTIONS ON SIGNAL PROCESSING (2006–2010).

He received the IEEE Signal Processing Magazine Best Column Award twice, in 2012 and 2014, the IEEE ComSoc Stephen O. Rice Prize in
Communications Theory in 2015, the IEEE ComSoc Leonard G. Abraham Prize in 2017, the IEEE ComSoc Best Tutorial Paper Award in 2018, and the IEEE ComSoc Fred W. Ellersick Prize in 2019.
\end{IEEEbiography}

\end{document}